\documentclass[review]{elsarticle}

\usepackage{hyperref}

\journal{Medical Image Analysis}

\usepackage{graphicx}
\usepackage{subfigure}
\usepackage{amssymb}
\usepackage{amsmath}
\usepackage{amsthm}
\usepackage[linesnumbered,ruled]{algorithm2e}
\usepackage{color}
\usepackage{mathrsfs}

\newdefinition{definition}{Definition}
\newproof{pf}{Proof}
\newproof{pot}{Proof of Theorem \ref{thm2}}
\usepackage{rotating}

\begin{document}

\begin{frontmatter}
\title{Recurrent convolutional neural networks for mandible segmentation from computed tomography}

\author[addr1,addr2,addr5]{Bingjiang Qiu}

\author[addr2,addr5]{Jiapan Guo\corref{mycorrespondingauthor}}
\cortext[mycorrespondingauthor]{Corresponding author}
\ead{j.guo@rug.nl}	

\author[addr1,addr3]{Joep Kraeima}

\author[addr1,addr3]{Haye H. Glas}

\author[addr4]{Ronald J. H. Borra}

\author[addr1,addr3]{Max J. H. Witjes}

\author[addr2,addr5]{Peter M. A. van Ooijen}

\address[addr1]{3D Lab\fnref{my3DLabfootnote}}
\address[addr2]{Department of Radiation Oncology}
\address[addr3]{Department of Oral and Maxillofacial Surgery}
\address[addr4]{Medical Imaging Center (MIC)}
\address[addr5]{Machine Learning Lab, Data Science Center in Health (DASH)\\ University Medical Center Groningen, University of Groningen, Hanzeplein 1, 9713GZ, Groningen, The Netherlands}


%
%
%

\begin{abstract}
Recently, accurate mandible segmentation in CT scans based on deep learning methods has attracted much attention. However, there still exist two major challenges, namely, metal artifacts among mandibles and large variations in shape or size among individuals. To address these two challenges, we propose a recurrent segmentation convolutional neural network (RSegCNN) that embeds segmentation convolutional neural network (SegCNN) into the recurrent neural network (RNN) for robust and accurate segmentation of the mandible. Such a design of the system takes into account the similarity and continuity of the mandible shapes captured in adjacent image slices in CT scans. The RSegCNN infers the mandible information based on the recurrent structure with the embedded encoder-decoder segmentation (SegCNN) components. 
The recurrent structure guides the system to exploit relevant and important information from adjacent slices, while the SegCNN component focuses on the mandible shapes from a single CT slice.
We conducted extensive experiments to evaluate the proposed RSegCNN on two head and neck CT datasets. The experimental results show that the RSegCNN is significantly better than the state-of-the-art models for accurate mandible segmentation.

\end{abstract}


\end{frontmatter}


\section{Introduction}

Globally, there are an estimated $ 650,000 $ new head and neck cancer cases and $ 330,000 $ deaths per year \cite{bray2018global}. Head and neck cancer is the 6th most incident cancer type globally \cite{weatherspoon2017oral}. Surgical tumor removal is an important curative treatment for this diseases. 
During surgical removal of malignant tumors in the oral cavity, a continuous resection of the jaw can be required. This resection, currently, is based on a 3D virtual surgical planning (VSP) \cite{kraeima2018multi} that enables accurate planning of the resection margin around the tumor, taking into account the surrounding jaw bone. 
Three dimensional VSP and accompanying 3D printed guides have been proven to be effective for resection and reconstruction of the mandible \cite{kraeima2018multi}. 
Such treatment planning approach has higher accuracy and certainty to guarantee sufficient margins for the removal of the tumor from the bone \cite{kraeima2018multi}. 
However, it requires manually intensive delineation of mandible organ \cite{kraeima2018multi}, which is very time-consuming and can delay treatment commencement. 
Thus, semiautomatic or automatic image segmentation would improve efficiency and reliability, as well as reduce the workload of technologists \cite{huff2018potential}. 

During the last decades, many researchers have investigated on the semiautomatic or automatic mandible segmentation in CT scans. 
Gollmer and Buzug \cite{gollmer2012fully} proposed a statictical shape model for mandible segmentation in 2012. A 3D gradient-based fuzzy connectedness algorithm for mandible segmentation was presented in 2017 \cite{torosdagli2017robust}. The study of Chuang et al. \cite{chuang2017novel} in 2017 applied a registration-based technique on semiautomatic mandible segmentation. Abdi et al. \cite{abdi2015automatic} proposed an automatic mandible segmentation approach using superior, inferior and exterior borders of mandibles in panoramic x-rays. A novel multi-atlas model was presented in \cite{chen2015multi}, which registered CT images with the atlases at the global level and allowed the fusion of multi-atlas-based segmentations and correlation-based label to be performed at the local level. 
Mannion-Haworth et al. \cite{mannion2015fully} was using active appearance models (AAM) built from manually segmented examples. High quality anatomical correspondences for the models are generated using a groupwise registration method. The models are then applied to segment ROIs in CT scans. 
An approach combined a multi-atlas segmentation with active shape model (ASM) was proposed by Albrecht et al. \cite{albrecht2015multi}
The above semiautomatic or automatic segmentation algorithms offer a potentially time-saving solution, the challenges in generalization ability and achieving expert performance still remain. 
The performances of these conventional methods are often affected by the noise or metal artifacts in the CT images, especially in the dental parts. Weak and false edges in the condyles often appear in the detected images, which frustrate the accurate segmentation of the mandible. 
The proposed deformable models, of which the parameters are determined according to the global characteristics of the target contour, are difficult to adapt to some local areas of the contour \cite{yuheng2017image, blaschke2004image, bankman2008handbook}. 

In recent years, deep learning based methods have proven better performance than traditional segmentation methods. In particular, several deep learning based approaches for mandible segmentation have been proposed \cite{ibragimov2017segmentation, zhu2018anatomynet, tong2018fully, qiu2019automatic}. 
Ibragimo et al. \cite{ibragimov2017segmentation} presented the first attempt of using CNN to segment organs-at-risks in head and neck CT scans. 
The AnatomyNet of Zhu et al. \cite{zhu2018anatomynet} in 2018 applied a 3D Unet architecture using residual blocks in encoding layers and a new loss function combining Dice score and focal loss in the training process. 
Tong et al. \cite{tong2018fully} proposed a fully CNN (FCNN) with shape representation model to segment organs-at-risks in CT scans. 
Despite the above mentioned CNN architectures have achieved promising results, challenges remain in developing clinical applicable algorithms. 

In this paper, our major contributions are in two folds. First, we propose to augment the classic segmentation networks structured as the encoder-decoder architecture with the recurrent neural networks that facilitate neuron connections to form a directed graph along a temporal (anatomical) sequence. Such a consideration allows segmentation inference based on the adjacent mandible regions. Second, the proposed framework enables 3D segmentation of the mandibles in CT scans in a way that reduces computational complexity of the model without loss of image quality. This is in contrary to the state-of-the-art (SOTA) 3-dimensional segmentation algorithms \cite{cciccek20163d}\cite{milletari2016Vnet} which usually down-sample the raw data to overcome memory issues.


\section{Related work} \label{relatedwork}

Convolutional neural network for semantic segmentation was initiated by Long et. al. who proposed the idea on fully convolutional network (FCN). It enabled an end-to-end image segmentation approach that is capable to cope with arbitrary sizes of input images \cite{long2015fully, garcia2017review}. Further based on FCN, Ronneberger et al. \cite{ronneberger2015Unet} published U-Net for medical image segmentation which has been so far one of the most popular image segmentation architecture. 
Later on, SegUnet \cite{kamal2019automatic} and attention Unet (AttUnet) \cite{oktay2018attention} are creative expansion techniques by taking into account the attention mechanism. 
In the following, we introduce briefly these networks.

\subsection{Unet} \label{seciton:unet}
\begin{figure}
	\centering
	\includegraphics[width=300pt]{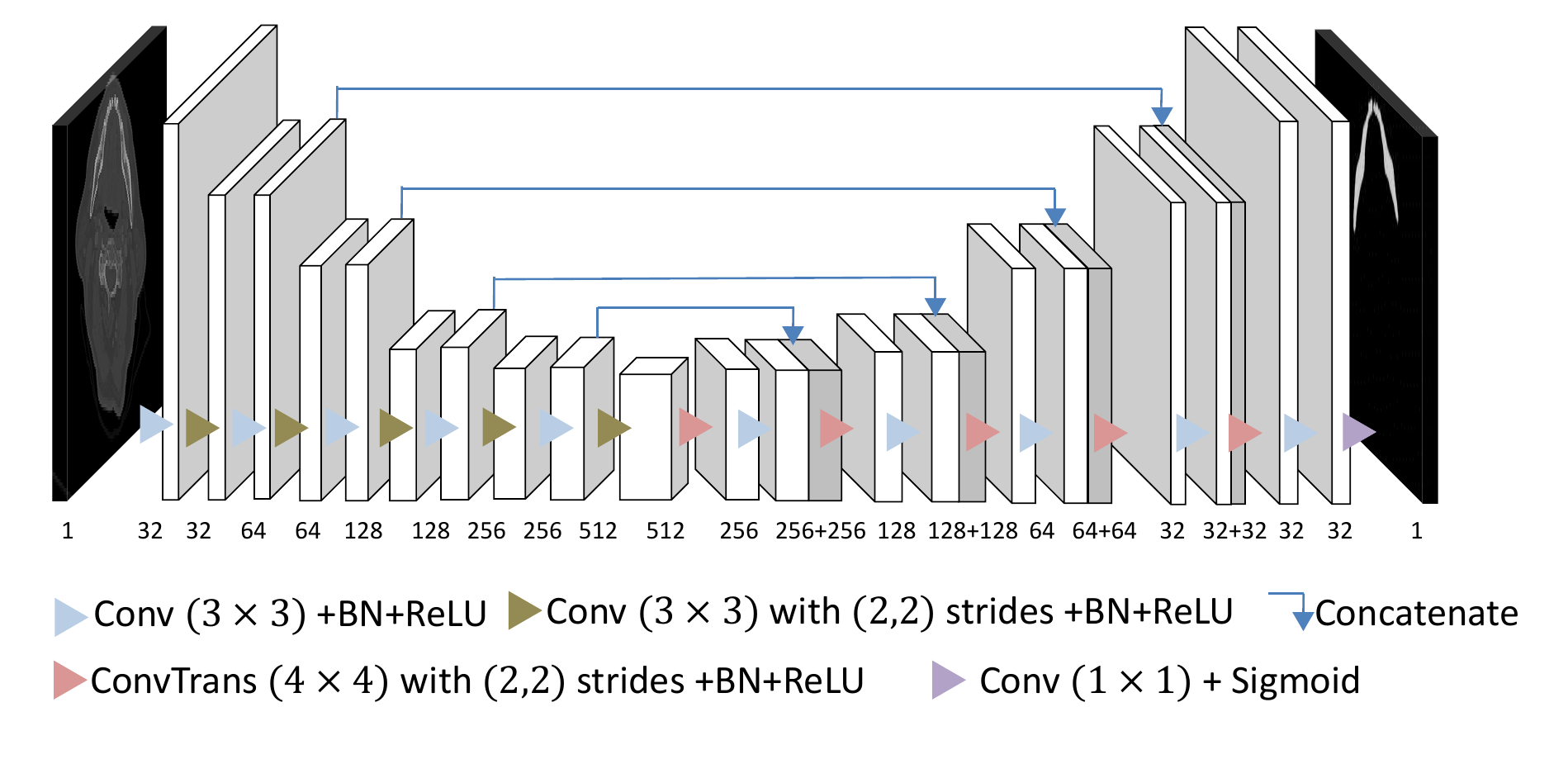}
	\caption{An overview of Unet. Input image is filtered step-wise and down-sampled by factor of 2 on each convolution with stride of (2,2) in the encoding path, while the feature map in the decoding path is up-sampled by factor of 2 on each transposed convolution with stride of (2,2). The final feature maps are fed to a convolution layer with size of $ (1 \times 1) $ and a Sigmoid function for pixel-wise prediction.}
	\label{fig:unet}
\end{figure}

In 2015, Ronneberger et al. \cite{ronneberger2015Unet} proposed Unet for biomedical image segmentation . 
The Unet consists of two paths. First path is the contraction path (also called encoder) which is used for feature extraction. It is a stack of several convolutional layers and max pooling layers. The second path is the up-sampling path (also known as decoder) which maps the feature representation back into the input data space. In every up-sampling step, feature maps are concatenated with those of the same resolution from the encoder path. It only contains convolutional layers and does not contain any dense layer in order to process input image of any size. 
For the implementation of this work, we use a modified Unet \cite{yan2018symmetric} which does not include the pooling and up-sampling layer. 
We employ the convolution kernel with stride of $ (2, 2) $ instead of max pooling while transposed convolution operator with stride of $(2, 2)$ instead of up-sampling. 
Figure \ref{fig:unet} shows the architecture of the implemented Unet. 

\subsection{SegUnet}\label{seciton:segunet}
Kamal et al. presented SegUnet \cite{kamal2019automatic} in 2019, which is motivated by Unet \cite{ronneberger2015Unet} and Segnet \cite{badrinarayanan2017segnet}.  
The network also consists of an encoder and a decoder. 
In the encoder stage, each convolution block consists of two convolution layers, BN and ReLU activation function. After each block, a max pooling operation is used for further enlarging the receptive fields. The index of the maximum element is used later during up-sampling. The number of feature channels is doubled at each down-sampling step. 
In the decoder stage, the feature map is upsampled using the pooling indices from the encoder process like Segnet  \cite{badrinarayanan2017segnet} and then concatenated with the corresponding feature map from the encoder like Unet shown in section \ref{seciton:unet}.

\subsection{AttUnet}
In order to segment small objects with large shape variability, Oktay  et al. \cite{oktay2018attention} proposed Attention Unet (AttUnet) by integrating attention gates (AG) in Unet. 
They proposed a novel attention gating module within skip concatenation of Unet, which allows attention coefficients to be more specific to local regions. It has the same encoder as SegUnet. In the decoder stage, each up-sampling block consists of up-sampling, convolution layer, BN and ReLU activation function and then concatenates with a feature map that is the pixel-wise multiplication of feature map and attention coefficients.

\subsection{Recurrent neural network (RNN)}
Recurrent neural network (RNN) is a class of neural networks formed with a self-connected hidden layer in order to explore correlations between sequential information \cite{graves2008novel}. 
RNN produces an output at each time step and has recurrent connections between hidden units that result in the sharing of parameters through a very deep computational graph \cite{goodfellow2016deep}. In this structure, recurrent networks can process sequences of varied lengths while traditional CNNs can only use a fixed length of input. Besides, it can make use of the context which remains in the network's internal state. RNN is widely applied in many tasks in natural language processing that deals with mostly sequential data \cite{young2018recent}. 

\section{Methods} \label{section:methods}
\begin{figure}
	\centering
	\includegraphics[width=350pt]{{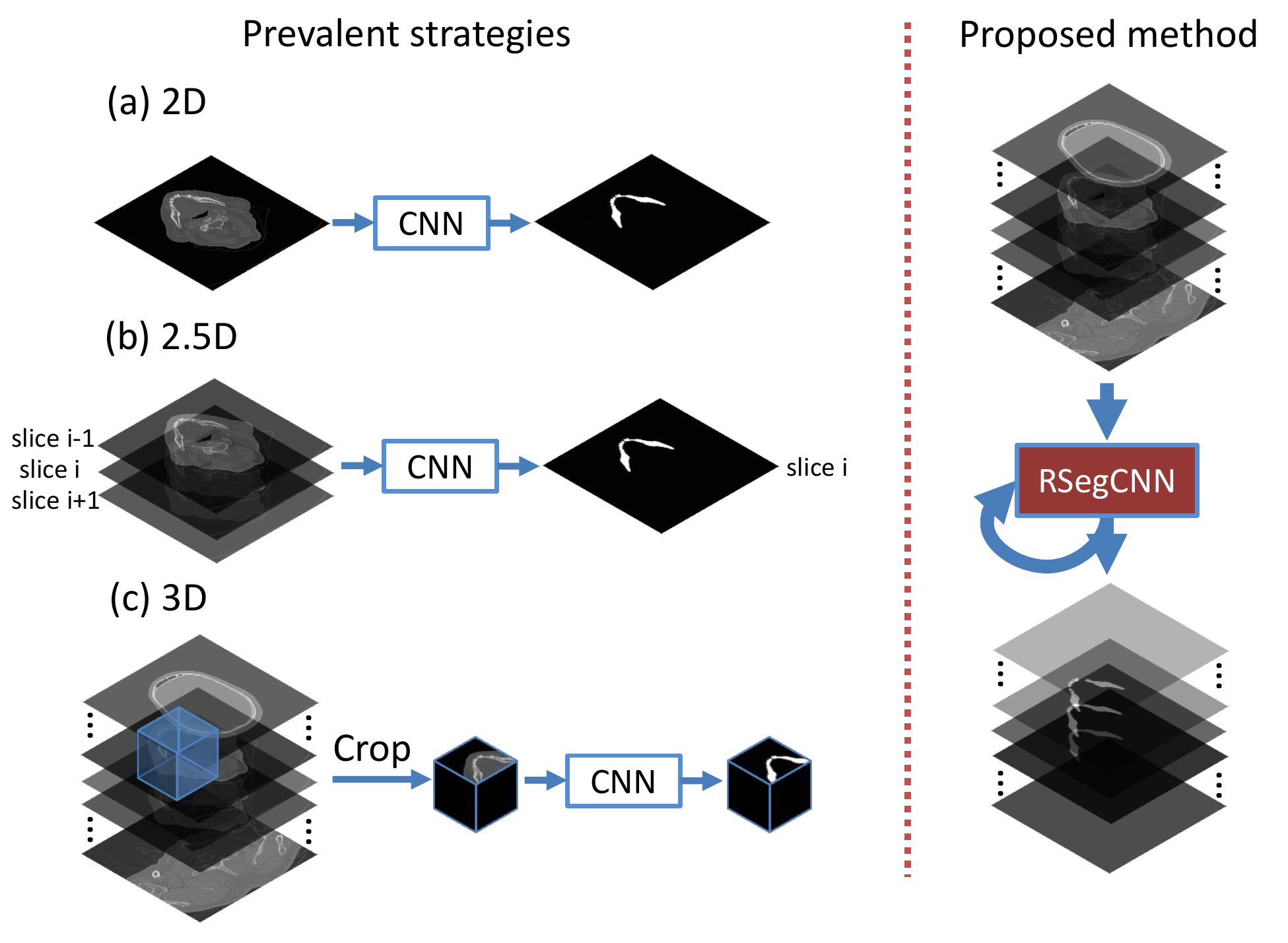}}
	\caption{Illustration of the prevalent strategies of performing segmentation of mandible. CNN indicates means the encoder-decoder based networks for mandible segmentation while RSegCNN represents the proposed model. The prevalent strategies are as follows. (a) The use of 2D segmentation network directly for slice by slice based mandible segmentation. (b)2.5D refers to using 2D convolution neural network with the input of adjacent slices from the volumetric images. (c)3D indicates the use of 3D CNN with the input of 3D volume cropped from the volumetric images. The proposed method integrates the recurrent unit with segmentation network into an end-to-end framework.  }
	\label{fig:prevalent_strategies_1}
\end{figure}

Encoder-decoder based deep learning segmentation architecture cannot take into consideration the continuity of anatomical structures in the organs. Therefore, we propose a novel approach that adopts the recurrent neural connection with the imbedding of the conventional encoder-decoder structure in order to perform 3D mandible segmentation. 
Figure \ref{fig:prevalent_strategies_1} illustrates three strategies to deal with 3D medical images in order to feed deep learning algorithms. 
An intuitive method is to train a 3D network to process volume data directly. 3D segmentation network can extract 3D structure information, but it still has two shortcomings: 
(i) cropping small 3D patches is frequently used to reduce the GPU memory, which can easily cause degradation in the segmentation task when the object spans several patches;  
(ii) it is still difficult and time-consuming to integrate patch output into the final prediction in testing. 
Thus, there are two alternative approaches to avoid using 3D networks which would occupy very high computational resource. The first method is to slice 3D volume into 2D images and train in the 2D network, which processes each slice independently like natural image segmentation. However, 2D network utilizes information from the entire 2D slice and lacks information about the relationship between slices (similarity and continuity of the organ). In order to learn anatomical information in 2D network, the concept of 2.5D was first used in Han's work \cite{han2017automatic}. It refers to using 2D networks with the input of adjacent slices from the volumetric images. Despite the promising performance of applying 2.5D strategy, 
the 2D networks using 2.5D data as input cannot fully utilize the 3D contexts, which limits the segmentation performance \cite{li2018h}. 
In this work, we propose a novel model for 3D image segmentation that combines recurrent module and encoder-decoder based segmentation networks, as shown in the right of Figure \ref{fig:prevalent_strategies_1}. This strategy utilizes 2D segmentation algorithms and also considers the anatomical structure continuity in a 3D form.  

\subsection{Recurrent segmentation CNN (RSegCNN)}

\begin{figure}
	\centering
	\includegraphics[width=350pt]{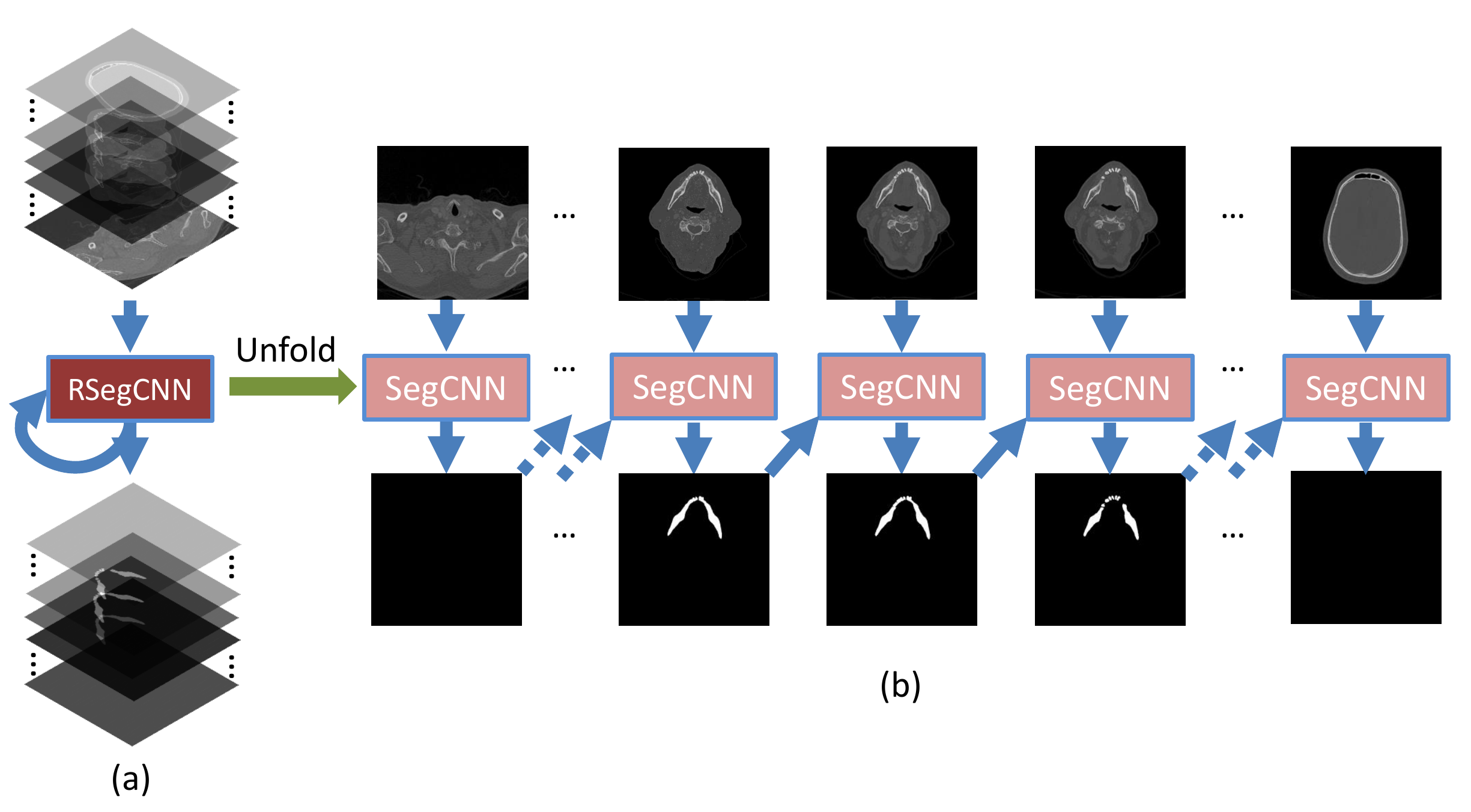}
	\caption{The overall graphic scheme of the proposed methods. The architecture of RSegCNN with two components: (a)The RSegCNN and its loss drawn with recurrent connections. (b) The same seen as a time-unfolded computational graph, where each node is now associated with one particular time instance. }
	\label{fig:RSegCNN}
\end{figure}

CNN is usually a feed-forward architecture, while recurrent connections are plentiful in the visual system  \cite{liang2015recurrent, abbott2001theoretical}. The feed-forward architectures were inspired by the biological neural networks in the brain \cite{kriegeskorte2015deep}. Recurrent connections are common in the neocortex, and recurrent synapses usually exceed feed-forward and feedback synapses \cite{liang2015recurrent, abbott2001theoretical}. 
Motivated by this fact, we propose a recurrent segmentation CNN strategy for mandible segmentation by incorporating the slice based segmentation algorithms into recurrent connections. 
The architecture is illustrated in Figure \ref{fig:RSegCNN}. 

Different from the typical recurrent convolutional neural network (RCNN) \cite{liang2015recurrent, lai2015recurrent} that uses recurrent connections within the same layer of the deep learning models, we use recurrent connections within the original whole segmentation networks. In such a way, the model is able to utilize the information obtained from the prediction of the previous adjacent CT slice.
We name the proposed approach RSegCNN which refers to recurrent convolutional neural networks for segmentation. The recurrent architecture consists of the composition of the conventional convolutional segmentation networks introduced in Section \ref{relatedwork}, in which the term SegCNN indicates any kind of CNN based image segmentation architecture, including Unet, SegUnet, AttUnet, but is not limited to these. Each unit is fed with the previous label predictions as shown in Figure \ref{fig:RSegCNN}. RSegCNN can process various lengths of slice sequences without any pre-processing in shape. 

\subsection{Problem definition} \label{section:methods_problem_definition}
For the simplicity of explanation, let $ X = \{x^1, ..., x^t, ..., x^n\}$ denotes the head and neck CT slices with the corresponding ground truth $ Y = \{y^1, ..., y^t, ..., y^n\} $ and the predicted maps $\hat{Y} = \{\hat{y}^1, ..., \hat{y}^t, ..., \hat{y}^n\}$, where $ n $ denotes the number of slices and $ t $ denotes the $ t $-th slice of the CT scan. Here $ t $ is equal to $ t $ time step in the implementation of RNN \cite{elman1990finding, goodfellow2016deep}. In this task, RSegCNN maps a sequence input slice to a sequence output of the same length, i.e., $ \hat{Y} = f(X) $. 
$ f(\cdot) $ denotes the learned target function that maps input to the ground truth. 
The output of the unfolded RSegCNN after $ t $ steps is represented as: 
\begin{equation}
\hat{y}^t = f(\hat{y}^{(t-1)}, x^t; \theta), 
\end{equation}
where $ \theta $ represents the network parameters. 

The RSegCNN model is trained by minimizing the error of its prediction and the ground truth. 
Two loss functions have been used during the training stage, namely, Dice and binary cross entropy (BCE) loss. These loss functions are selected due to their potiential to deal with imbalanced data.  
\begin{equation}
\label{eq:loss}
\mathcal{L} = \omega_1 \times \mathcal{L}_{BCE} +  \omega_2 \times \mathcal{L}_{Dice}, 
\end{equation}
where $ \omega_1 $ and $ \omega_2 $ control the amount of BCE term and Dice term contribution in the loss function $ \mathcal{L} $. $\mathcal{L}_{BCE}$ and $\mathcal{L}_{Dice}$ are defined as follows: 
\begin{equation}
\label{eq:loss_BCE}
\mathcal{L}_{BCE}(\hat{y}, y) =- \frac{1}{N}\sum_{i=1}^{N} y_i\log(\hat{y}_i)+(1-y_i)\log(1-\hat{y}_i), 
\end{equation}
\begin{equation}
\label{eq:loss_Dice}
\mathcal{L}_{Dice}(\hat{y}, y) = 1 - \frac{2 \sum_{i=1}^{N} y_i \hat{y}_i }{\sum_{i=1}^{N} y_i + \hat{y}_i}. 
\end{equation}
Here, $y_i$ and $\hat{y}_i $ represent the ground truth and the predicted probability of pixel $ i $ respectively, $ N $ is the number of pixels. 

According to equation (\ref{eq:loss_BCE}), the term $ (1-y_i)\log(1-\hat{y}_i)$ penalizes false positives (FPs) as it is $ 0 $ when the prediction probability is correct, and $y_i\log(\hat{y}_i) $ penalizes false negatives(FNs) \cite{taghanaki2019combo}. Therefore, the BCE term is able to control the trade-off between FPs and FNs in the pixel-wise segmentation task. In spite of that, the networks with only BCE as loss function are often prone to generate more false positives in the works of segmentation \cite{abulnaga2018ischemic}. Sudre et al. \cite{sudre2017generalised} have proven that Dice loss gives better performance for one-target segmentation and has capacity of predicting the fine appearance features of the object. Dice loss is based on the metric Dice coefficient that measures the proportion of overlapping between the resulting segmentation and the ground truth. Thus the combination of loss functions can control the penalization of both FPs and FNs by BCE term, and drag the model parameters out of local minima via Dice term, simultaneously. 

The training procedure of RSegCNN is the same as the traditional CNN, where the trainable weights are updated with the backpropagation through time (BPTT) algorithm \cite{lecun2015deep}. 
According to equation (\ref{eq:loss}), the loss for $ t $-th step with prediction $ \hat{y} $ with respect to ground truth $ y $ is
\begin{equation}
\mathcal{L}^t(\hat{y}^t, y^t) = \omega_1 \times \mathcal{L}_{BCE}(\hat{y}^t, y^t) +  \omega_2 \times \mathcal{L}_{Dice}(\hat{y}^t, y^t), 
\end{equation}
in which each $ \mathcal{L}^t $ used only at step $ t $. 
Then the total loss for a given sequence of slices $ X = \{x^1, ..., x^t, ..., x^n\} $  paired with a sequence of ground truth $ Y = \{y^1, ..., y^t, ..., y^n\}  $ is the sum of the losses over all the steps. 
\begin{equation}\label{eq:total_loss}
\begin{split}
\mathcal{L}(\{\hat{y}^1, ..., \hat{y}^n\},\{y^1, ..., y^n\}) =& \sum_{t=1}^{n} \mathcal{L}^t(\hat{y}^t, y^t) \\
=& \sum_{t=1}^{n} \omega_1 \times \mathcal{L}_{BCE}(\hat{y}^t, y^t) +  \omega_2 \times \mathcal{L}_{Dice}(\hat{y}^t, y^t)
\end{split}
\end{equation}

Here we use BPTT to optimize the model parameters $\theta$ to minimize the loss $ \mathcal{L} $ based on the chain rule \cite{magin2006fractional}. We start recursively from the last node before the final loss:  
\begin{equation}\label{eq:partial_total_loss_t}
\frac{\partial \mathcal{L}}{\partial \mathcal{L}^t} = 1. 
\end{equation}

The gradient $ \frac{\partial \mathcal{L}^t}{\partial \hat{y}^t_j} $ on the outputs at step $ t $, for all $ j, t $, is as follows: 
\begin{equation}\label{eq:partial_probability_t}
\begin{split}
\frac{\partial \mathcal{L}^t}{\partial \hat{y}^t_j}  =& \omega_1 \frac{ \partial\mathcal{L}_{BCE}(\hat{y}^t, y^t) }{\partial \hat{y}^t_j} + \omega_2 \frac{\partial \mathcal{L}_{Dice}(\hat{y}^t, y^t)}{\partial \hat{y}^t_j} \\
=& -\frac{\omega_1}{N} \left( \frac{y^t_j}{\hat{y}^t_j} - \frac{1-y^t_j}{1-\hat{y}^t_j}\right) - \omega_2 \frac{2y^t_j}{ \sum_{i=1}^{N} y^t_i + \hat{y}^t_i } + \\
&  \omega_2 \frac{2 \sum_{i=1}^{N} y^t_i \hat{y}^t_i }{\left( \sum_{i=1}^{N} y^t_i + \hat{y}^t_i \right)^2 } 
\end{split}
\end{equation}

The sigmoid function $ \sigma = \frac{1}{1+\exp (-x)}$ is used as the activation function in the last layer for all the basic segmentation networks as illustrated in Section \ref{relatedwork}. We denote by $ o^t $ the output from the last layer of the network and then use the sigmoid function to obtain the probabilities $ \hat{y}^t $ over the outputs, i.e. $ \hat{y}^t_j=\sigma(o^t_j)= \frac{1}{1+\exp (-o^t_j)} $, for all $ j, t $. So, the derivative of probabilities $ \hat{y}^t $ with respect to $ o^t_j $ is
\begin{equation}\label{eq:partial_o_t}
\begin{split}
\frac{\partial \hat{y}^t_j }{\partial o^t_j} =& \frac{\partial \frac{1}{1+\exp (-o^t_j)}}{\partial o^t_j} \\
=& \sigma(o^t_j) (1-\sigma(o^t_j))\\ 
=& \hat{y}^t_j (1-\hat{y}^t_j) 
\end{split}
\end{equation}

We then iterate backward in steps using BPTT, starting from $ t = n $ (the end of the sequence slice) down to $ t = 1 $(the beginning of the suquence slice). We calculate the gradient of $ \mathcal{L} $ with respect to  each weight $ \theta_j $ using the above equations (\ref{eq:total_loss})-(\ref{eq:partial_o_t}): 
\begin{equation}
\left. \frac{\partial \mathcal{L}}{\partial \theta^t_j}\right|_{\{x_1,...,x_t;y_1, ..., y_t\}}    = \left.\frac{\partial \mathcal{L}}{\partial \mathcal{L}^t}\frac{\partial \mathcal{L}^t}{\partial \hat{y}^t_j}\frac{\partial \hat{y}^t_j }{\partial o^t_j}\frac{\partial o^t_j}{\partial \theta^t_j}\right|_{\{x_1,...,x_t;y_1, ..., y_t\}} 
\end{equation}
By recursively computing the gradient of $ \mathcal{L} $ with respect to $ \theta_j $, we can compute the gradients for all parameters in RSegCNN. Furthermore, we update the model parameters. 
The parameter updates are achieved with Adam \cite{kingma2014adam} algorithm with a learning rate $ r $:
\begin{equation}
\theta^t_j \leftarrow \theta^t_j + \left. \lambda_{Adam}(r) \times \frac{\partial \mathcal{L}}{\partial \theta^t_j}\right|_{\{x_1,...,x_t;y_1, ..., y_t\}} 
\end{equation}

\section{Experiments and Results}\label{section:experiments}
A series of experiments are carried out to evaluate segmentation performance of the proposed strategy. To evaluate our method, two datasets are used in which our own dataset is obtained from University Medical Center Groningen (UMCG) and a public dataset is collected by Public Domain Database for Computational Anatomy (PDDCA) version 1.4.1 \cite{raudaschl2017evaluation}. In addition, the experimental results of our method are presented and compared with other state-of-the-art methods on the PDDCA dataset.

\subsection{Experimental Setup}

\subsubsection{Implementation details}

We implement the approach proposed in Section \ref{section:methods} by using Pytorch \cite{paszke2017automatic} package developed by Facebook. 
The CNN models are trained on a workstation equipped with Nvidia GPU K40m of 12GB memory. The weights of the BCE loss term $ \omega_1 $ and the Dice loss term $ \omega_2 $ in the loss function are both set to $ 0.5 $. We use Adam optimization with a learning rate of $ r = 10^{-4} $. We set the total number of epochs to 40 and 80 for UMCG dataset and PDDCA dataset, respectively. 
Moreover, an early stopping strategy is utilized if there is no improvement in the validation set after 10 epochs in order to avoid over-fitting.

\subsubsection{Evaluation metrics} 
For quantitative analysis of the experimental results, several performance metrics are considered, including Dice coefficient (Dice), average symmetric surface distance(ASD) and $ 95\% $ Hausdorff distance (95HD). 

Dice coefficient is often used to measure consistency between two objects \cite{ghafoorian2017location}. Therefore, it is widely applied as a metric to evaluate the performance of image segmentation algorithms. It is defined as: 
\begin{equation}
\label{eq:Dice}
{\rm Dice} = \frac{2 \sum_{i=1}^{N} y_i \hat{y}_i }{\sum_{i=1}^{N} y_i + \hat{y}_i},  
\end{equation}
where 
$y_i$ and $ \hat{y}_i $ represent the ground truth and the predicted probability of pixel $ i $ respectively, $ N $ is the number of pixels.

The average symmetric surface distance (ASD) \cite{tong2018fully} computes the average distance between the boundaries of two object regions. It is defined as: 
\begin{equation}
\label{eq:ASD}
{\rm ASD}(A,B) = \frac{d(A, B)+d(B, A)}{2}, 
\end{equation}

\begin{equation}
\label{eq:ASD_}
d(A,B) = \frac{1}{N}\sum_{a \in A} \min_{b \in B} \|a-b\|, 
\end{equation}
where $\|.\|$ is the $ L_2 $ norm. $ a $ and $ b $ are corresponding points on the boundary of $ A $ and $ B $. 

Hausdorff distance (HD) measures the maximum distance of a point in a set $ A $ to the nearest point in the other set $ B $. It is defined as:
\begin{equation}
\label{eq:HD}
{\rm HD}(A,B) = \max (h(A, B), h(B, A))
\end{equation}
\begin{equation}
\label{eq:HD_}
h(A,B) = \max_{a \in A} \min_{b \in B} \|a-b\|
\end{equation}
where $h(A,B)$ is often called the directed HD. 
The maximum HD is sensitive to contours. 
When the image is contaminated by noise or occluded, the original Hausdorff distance is prone to mismatch \cite{raudaschl2017evaluation, taha2015metrics}. Thus, Huttenlocher proposed the concept of partial Hausdorff distance \cite{huttenlocher1992comparing}.
The 95HD is similar to maximum HD and selects $ 95\% $ closest points in set $ B $ to the point in set $ A $ in equation (\ref{eq:HD_}) to calculate $ h(A,B) $: 
\begin{equation}
\label{eq:95HD}
95{\rm HD} =  \max (h^{95\%}(A, B), h^{95\%}(B, A))
\end{equation}
\begin{equation}
\label{eq:95HD_}
h^{95\%}(A,B) = \max_{a \in A} \min_{b \in B^{95\%}} \|a-b\|
\end{equation}
The purpose for using this metric is to eliminate the impact of a very small subset of inaccurate segmentations on the evaluation of the overall segmentation quality. 

\subsection{The UMCG H$\&$N Dataset} 
\subsubsection{Dataset preparation}

The dataset from UMCG contains 109 CT scans reconstructed with a kernel of Br64, I70h(s) or B70s of Siemens. Each scan consists of $221$ to $955$ slices with a size of $512 \times 512$ pixels. The pixel spacing varies from 0.35 to 0.66 mm and the slice thickness varies from 0.6 to 0.75 mm. The corresponding mandible segmentation labels were obtained by an experienced researcher using Mimics software version 20.0 (Materialise, Leuven, Belgium), and confirmed by a trained clinician.  

\subsubsection{Results}
\begin{sidewaysfigure}
	\includegraphics[width=\columnwidth]{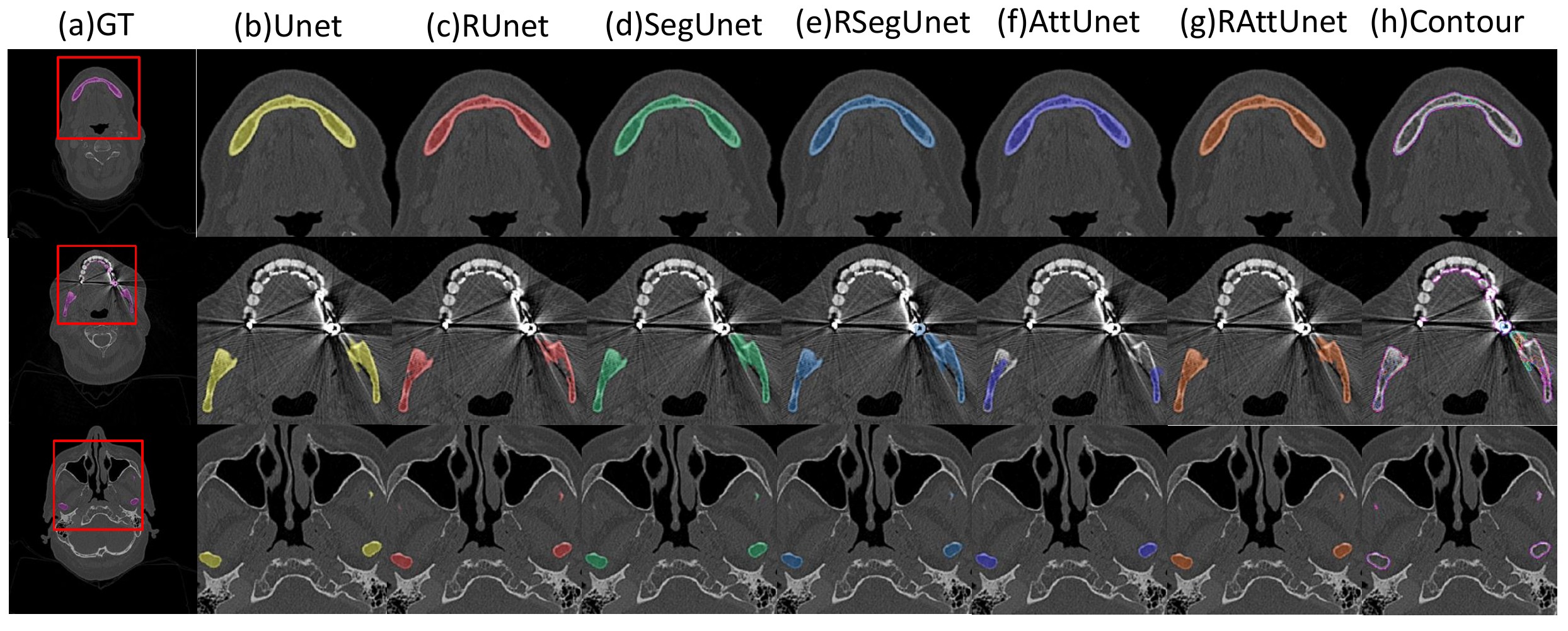}%
	\caption{Examples of the automatic segmentation of mandibles in UMCG dataset.}
	\label{fig:2D_segmentation_results}
\end{sidewaysfigure}

\begin{figure}
	\centering
	\includegraphics[width=350pt]{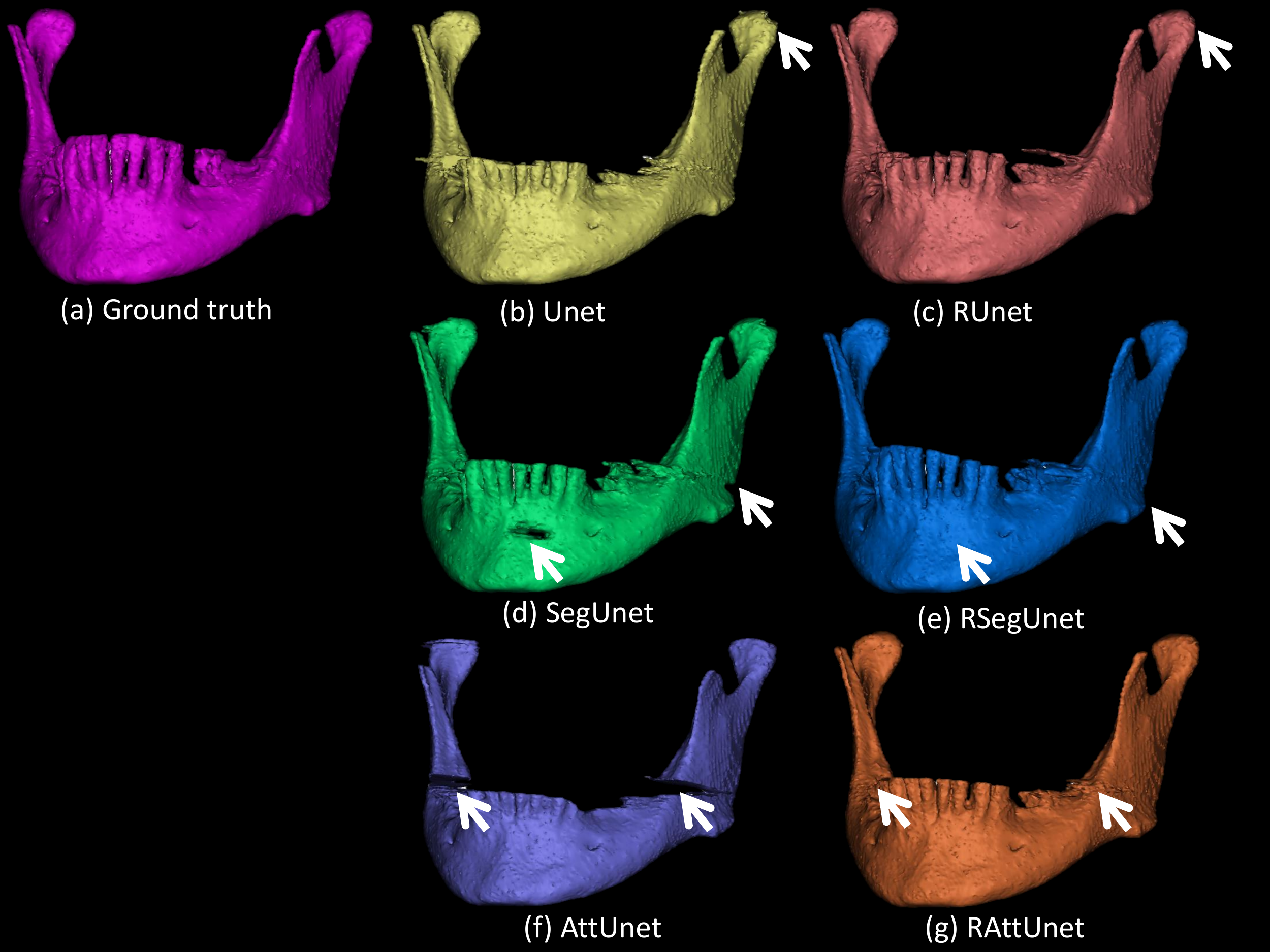}
	\caption{3D view of a case from the UMCG dataset. From (a) to (g): Ground truth, Unet, RUnet, SegUnet, RSegUnet, AttUnet and RAttUnet. }
	\label{fig:3D_view_of_UMCG_dataset_pic}
\end{figure}

We first compare the proposed methods with and without RNN unit. For brevity, we refer to our methods as RUnet, RAttUnet and RSegUnet, that use Unet, AttUnet and SegUnet as the base unit of the proposed RSegCNN, respectively. 
We randomly chose 90 cases as training, 2 cases as validation and 17 cases as test. 
The training takes approximately 40 hours while the test on one scan is about 1.5 minutes. 
Figure \ref{fig:2D_segmentation_results} illustrates several examples resulting from the proposed approach with and without RNN unit on the test scan. 
Figure \ref{fig:3D_view_of_UMCG_dataset_pic} shows a case of the automatic segmentation in the 3D view of Unet, RUnet, SegUnet, RSegUnet, AttUnet and RAttUnet. 
The examples shown in Figure \ref{fig:3D_view_of_UMCG_dataset_pic} (b-c) indicate that the proposed approach incorporated with RNN module is more robust for the segmentation of condyles (pointed by the white arrows) in mandible that usually appear with faint boundaries. 
The examples shown in Figure \ref{fig:3D_view_of_UMCG_dataset_pic} (d-e) show that the traditional approach is easier to lead to a wrong segmentation in the body and angle of the mandible (pointed by the white arrows) while the proposed strategy can make a smooth segmentation results since it makes use of the previous information obtained from the last step. 
The examples shown in Figure \ref{fig:3D_view_of_UMCG_dataset_pic} (f-g) demonstrate that RSegCNN can also handle the weak and thin object, i.e., ramus of the mandible (pointed by the white arrows). 
The conventional methods usually lead to a wrong segmentation within the whole mandible as shown Figure \ref{fig:3D_view_of_UMCG_dataset_pic}. 
The visual comparison of the automatic segmentation with manual segmentation demonstrates the effectiveness of our method on automatic segmentation of mandible.  

To quantitatively compare the proposed approach with other methods, we compute the Dice scores, ASD and 95HD values of the six methods. Table \ref{table:umcg} lists Dice, ASD and 95HD as well as the corresponding standard deviation. 
In general, the average values of these metrics obtained from our proposed method are better than the other methods. 
From the Table, it can be observed that our method gives smaller mean errors. As shown in Table \ref{table:umcg}, the proposed method RSegCNN outperforms the conventional methods. 
Figure \ref{fig:2D_segmentation_results}, Table \ref{table:umcg} and Figure \ref{fig:3D_view_of_UMCG_dataset_pic} indicate that the proposed approach is quite accurate in segmenting mandible, especially, condyles and ramus of the mandible. These significant improvements in segmentation indicate that RSegCNNs help improve the existing encoder-decoder based segmentation approaches since they take the similarity of the neighborhood slices into consideration.

\begin{table}[]
	\centering
	\caption{Quantitative comparison of segmentation performance in UMCG dataset between the proposed method and that without the implementation of RNN module.}%
	\label{table:umcg}
	\begin{tabular}{|l|c|c|c|c|c|c|}
		\hline
		& Dice(\%)    & ASD(mm) & 95HD(mm) \\ \hline
		Unet                                          & 95.95($\pm$2.24)     & 0.3615($\pm$0.3366)	       & 4.0145($\pm$4.6487)  \\ \hline
		RUnet                                         &\textbf{ 97.53($\pm$1.65)} & \textbf{0.207($\pm$0.2623)} & \textbf{2.3975($\pm$4.6051)}  \\ \hline
		SegUnet                                       & 96.3($\pm$2.06) & 0.2794($\pm$0.2447) & 3.7958($\pm$4.3662)   \\ \hline
		RSegUnet                              & \textbf{97.48($\pm$1.70)} & \textbf{0.2170($\pm$0.3491)} & \textbf{2.6562($\pm$5.7014)}   \\ \hline
		AttUnet                                       & 94.21($\pm$3.34) & 0.6929($\pm$0.837) & 5.1368($\pm$3.2194)   \\ \hline
		RAttUnet                                      & \textbf{96.57($\pm$1.69)} & \textbf{0.2978($\pm$0.234)} & \textbf{2.4068($\pm$1.5479)}   \\ \hline
	\end{tabular}
\end{table}

\subsection{PDDCA dataset}
\subsubsection{Dataset preparation}
We also test the proposed strategy on the public dataset PDDCA \cite{raudaschl2017evaluation}. This dataset contains 48 patient CT scans from the Radiation Therapy Oncology Group (RTOG) 0522 study, a multi-institutional clinical trial, together with manual segmentation of left and right parotid glands, brainstem, optic chiasm and mandible. Each scan consists of $76$ to $360$ slices with a size of $512 \times 512$ pixels. The pixel spacing varies from 0.76 to 1.27mm, and the slice thickness varies from 1.25 to 3.0 mm. According to the Challenge description, we follow the same training and testing protocol \cite{raudaschl2017evaluation}. Forty out of the 48 patients in PDDCA with manual mandible annotations are used in this study \cite{raudaschl2017evaluation, ren2018interleaved} in which the dataset is split into the training and test subsets, each with 25  (0522c0001-0522c0328) and 15 (0522c0555-0522c0878) cases, respectively. \cite{raudaschl2017evaluation}. 

\subsubsection{Results}
\begin{sidewaysfigure}
	\includegraphics[width=\columnwidth]{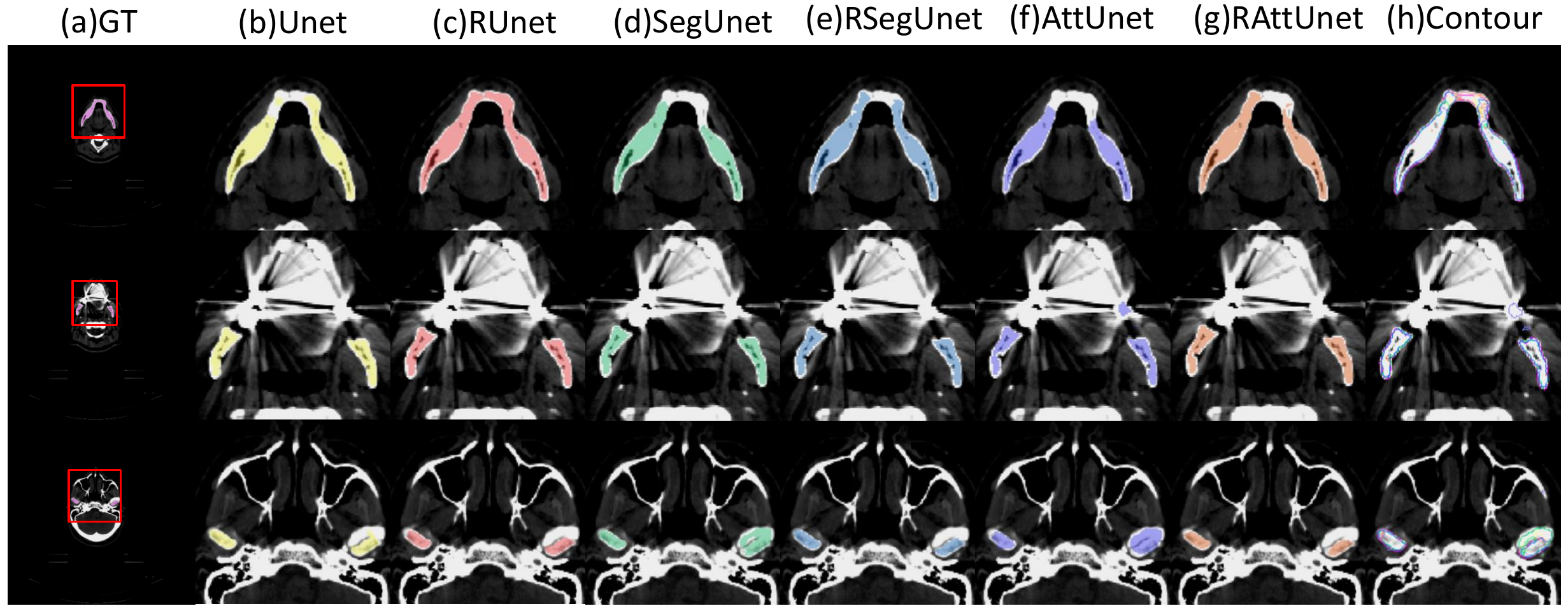}%
	\caption{Examples of the automatic segmentation of mandibles in PDDCA dataset.}
	\label{fig:2D_segmentation_results_pddca}
\end{sidewaysfigure}

Here the proposed methods are fine-tuned in the pre-trained models obtained based on the UMCG dataset. Figure \ref{fig:2D_segmentation_results_pddca} shows some examples of GT, Unet, RUnet, SegUnet, RSegUnet, AttUnet, RAttUnet, and the corresponding contours of the above methods. 
The segmentation results of the six methods on three different slices are shown in Figure \ref{fig:2D_segmentation_results_pddca}. 
As shown in the first row of Figure \ref{fig:2D_segmentation_results_pddca}, the encoder-decoder based models (e.g. Unet, SegUnet, AttUnet) fail to obtain satisfactory results in the main body of mandible. 
While the automatic segmentation results from our approach are much better. The second and the third rows in Figure \ref{fig:2D_segmentation_results_pddca} show better performances on the ramus and condyles of mandibles which are corrupted by noise and metal artifacts in its collection. In addition, the proposed strategy can deal with the condyles area that seems to be connected with the other bone structure, while the traditional methods cause boundary leakage, as shown in the last row of Figure \ref{fig:2D_segmentation_results_pddca}. 

Similarly, Table \ref{tab:PDDCA} shows a quantitative evaluation based on average Dice, ASD and 95HD used in the challenge \cite{raudaschl2017evaluation, tong2018fully}. 
Similar observation can be made, Table \ref{tab:PDDCA} indicates that our method leads to a significant improvement with respect to the metrics of Dice, ASD and 95HD. 

\begin{table}[]
	\centering
	\caption{Quantitative comparison of segmentation performance in PDDCA dataset of between the proposed method and without RNN module. We mark in bold the best performance in each metrics.}%
	\begin{tabular}{|l|c|c|c|c|c|c|}
		\hline
		& Dice(\%)      & ASD(mm) & 95HD(mm)  \\ \hline
		Unet                        & 94.15($\pm$1.31)  & 0.1827($\pm$0.0915)   & 2.0547($\pm$1.4431)  \\ \hline		
		RUnet                         & 94.71($\pm$1.35)  & \textbf{0.1353($\pm$0.0614)}   & 1.4098($\pm$0.8573)  \\ \hline	
		SegUnet                   & 94.69($\pm$1.33)  & 0.1765($\pm$0.0671)   & 1.5067($\pm$0.6938)  \\ \hline		
		RSegUnet                    & \textbf{95.10($\pm$1.21)} & 0.1367($\pm$0.0382)  & \textbf{1.356($\pm$0.4487)}  \\ \hline
		AttUnet                       & 92.99($\pm$1.25)  & 0.2924($\pm$0.2523)   & 3.1848($\pm$4.0571)  \\ \hline			
		RAttUnet                  & 93.87($\pm$1.29) & 0.1773($\pm$0.0515)  & 1.6397($\pm$0.6219) \\ \hline
	\end{tabular}
	\label{tab:PDDCA}
\end{table}

\begin{figure}
	\centering
	\includegraphics[width=350pt]{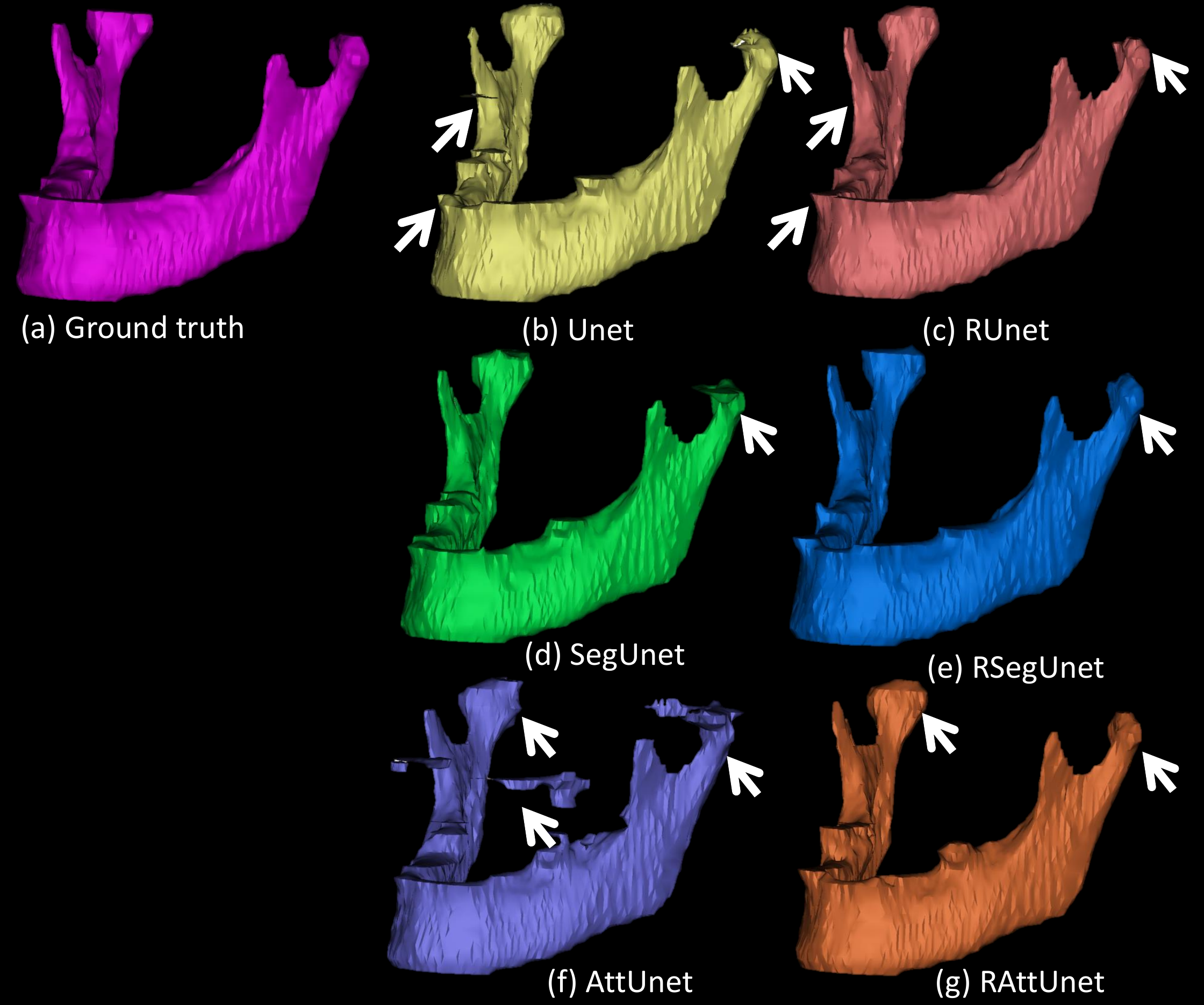}
	\caption{3D view of a case from of PDDCA dataset. From (a) to (g): Ground truth, Unet, RUnet, SegUnet, RSegUnet, AttUnet and RAttUnet.}
	\label{fig:3D_view_of_PDDCA_dataset_pic}
\end{figure}

Figure \ref{fig:3D_view_of_PDDCA_dataset_pic} also shows a case of the automatic segmentation in the 3D view of the six methods. 
Figure \ref{fig:3D_view_of_PDDCA_dataset_pic} demonstrates the models without RNN module are less effective in dealing with the scans with weak boundaries of condyles in mandible.   
The examples shown in Figure \ref{fig:3D_view_of_PDDCA_dataset_pic} (b)-(c) and (f)-(g) show that the encoder-decoder based methods can easily lead to an over-estimation for the upper teeth (pointed by the white arrows) in their CT images. In contrast, the proposed strategy can make a promising segmentation results since it makes use of the previous information obtained from its neighborhood slices. 
Our proposed approach based on the RNN module is able to segment the details of the mandibles more accurate.
Table \ref{tab:PDDCA} and Figure \ref{fig:3D_view_of_PDDCA_dataset_pic} demonstrate that RSegCNN achieves higher segmentation and quantification accuracy comparing to the existing methods. 

\subsubsection{Comparison with state-of-the-art methods}
As illustrated in Table \ref{tab:PDDCA_compare_with_papers}, the proposed segmentation methods are compared with several SOTA methods on Dice, ASD and 95HD. 
We can see that the methods with RNN unit outperfoms the existing approaches in PDDCA based on the metrics. 
In addition, our RSegCNN algorithm is able to segment more detailed information, especially in the weak and confusing boundaries such as the areas of condyles and ramus. 
That is because the architecture of RSegCNN is designed to consider the continuity of upper and lower slice. The conventional CNN frameworks fail to segment accurately condyles or ramus target of mandible, which can cause the ineffective automatic segmentation for 3D virtual surgical planning.

\begin{table}[]
	\centering
	\caption{Comparison of segmentation accuracy between the state-of-the-art methods and our method, bold fonts indicate the best performer for that measurement.}%
	\begin{tabular}{|l|c|c|c|c|c|c|}
		\hline
		& Dice(\%)      & ASD(mm) & 95HD(mm)  \\ \hline
		Multi-atlas\cite{chen2015multi}      & 91.7($\pm$2.34)   & -        & 2.4887($\pm$0.7610)    \\ \hline
		AAM\cite{mannion2015fully}         & 92.67($\pm$1)    & -        & 1.9767($\pm$0.5945)    \\ \hline
		ASM\cite{albrecht2015multi}             &88.13($\pm$5.55)    & -        & 2.832($\pm$1.1772)    \\ \hline
		CNN\cite{ibragimov2017segmentation}           & 89.5($\pm$3.6)   & -        & -       \\ \hline
		NLGM \cite{orbes2015head}         & 93.08($\pm$2.36)   & -        & -       \\ \hline
		AnatomyNet\cite{zhu2018anatomynet}        & 92.51($\pm$2)  & -        & 6.28($\pm$2.21)    \\ \hline
		FCNN\cite{tong2018fully}          & 92.07($\pm$1.15)  & 0.51($\pm$0.12)     & 2.01($\pm$0.83)    \\ \hline
		FCNN+SRM\cite{tong2018fully}      & 93.6($\pm$1.21)   & 0.371($\pm$0.11)     & 1.5($\pm$0.32)     \\ \hline
		CNN+BD\cite{kodym2018segmentation}  & \textbf{94.6($\pm$0.7)}   & 0.29($\pm$0.03)     & -       \\ \hline
		HVR\cite{wang2017hierarchical} & 94.4($\pm$ 1.3)   & 0.43($\pm$ 0.12)      & -       \\ \hline
		Cascade 3D-Unet\cite{wang2018organ}  & 93($\pm$1.9)   & -     & \textbf{1.26($\pm$0.5)}       \\ \hline
		RUnet                         & \textbf{94.71($\pm$1.35)}  & \textbf{0.1353($\pm$0.0614)}   & \textbf{1.4098($\pm$0.8573)}  \\ \hline			RSegUnet                    & \textbf{95.10($\pm$1.21)} & \textbf{0.1367($\pm$0.0382)}  & \textbf{1.356($\pm$0.4487)}  \\ \hline        RAttUnet                  & 93.87($\pm$1.29) & \textbf{0.1773($\pm$0.0515)}  & 1.6397($\pm$0.6219) \\ \hline
	\end{tabular}
	\label{tab:PDDCA_compare_with_papers}
\end{table}

\section{Discussion}
In this paper, we present a robust end-to-end method for mandible segmentation that combines the recurrent module and the encoder-decoder segmentation networks. Unlike other methods, the proposed approach utilizes 2D segmentation architecture that is embeded in the recurrent unit to acquire anatomical structure continuity in 3D form but also considers the mandible structure continuity in 3D form. 
Quantitative evaluation results shown in Table \ref{table:umcg}-\ref{tab:PDDCA} demonstrate that our proposed approach outperforms the conventional methods for mandible segmentation.
Besides, qualitative visual inspection in Figure \ref{fig:3D_view_of_UMCG_dataset_pic}-\ref{fig:3D_view_of_PDDCA_dataset_pic} illustrates that our automatic segmentation approach performs quite good comparing to the ground truth. 
The direct comparison (Table \ref{tab:PDDCA_compare_with_papers}) between with and without RNN module in PDDCA dataset illustrates that RSegCNN strategy significantly improves segmentation of mandible. 
Remarkably, we found that this kind of segmentation architecture is very robust for the weak and blurry edge segmentation. For instance, the networks can segment condyles or ramus of mandible quite well, both of which are very important regions for the 3D VSP.

The experimental results show that the proposed strategy is feasible and effective in 3D mandible segmentation, and can also be applied to other segmentation tasks. This strategy takes advantage of shape prior information, which considers the segmentation result of the previous slice as the shape prior to segment the current slice. Therefore, the proposed approach can help to learn the continuous structure of the mandible in 2D segmentation networks. In addition, the proposed approach utilizes RNN module that helps the extraction of spatial information of object based on the collection of context and shape information. This strategy can support further research on the 3D image segmentation. It can also help alleviate memory issues for 3D medical image segmentation as well as handle 3D segmentation tasks. 

In this study, we use 109 H\&N CT dataset and a small public dataset to validate our method, which cannot satisfactorily represent the average population in clinical practice. In practical automatic segmentation simulation, more data from different regions should be further explored. In addition, most of CT scans from the datasets exclude metal implants or dental braces that often make the CT scans noisy and blurry, and also make the segmentation network difficult to train. In our further study, more validation and evaluation will be performed to determine whether our present strategy is effective for real clinical practice.

\section{Conclusion} 
We propose an end-to-end approach for accurate segmentation of the mandible from H\&N CT scans. Our approach incorporates the encoder-decoder based segmentation algorithms into recurrent connections and uses a combination of Dice and BCE as loss function. We implement the proposed approach on 109 H\&N CT scans from our dataset and 40 scans from PDDCA public dataset. The experimental results have shown that the RSegCNN strategy can yield more significant performance than the conventional algorithms in terms of quantitative and qualitative evaluation. The proposed RSegCNN has potential for automatic mandible segmentation by learning spatial structured information.

\section*{Conflict of interest}
The authors declare no conflict of interest.

\section*{Acknowledgments}
The author is supported by a joint PhD fellowship from China Scholarship Council (CSC 201708440222). The authors acknowledge Erhan Saatcioglu for the training data preparation. 

\bibliography{MIDL_using_nips_2017}

\end{document}